\documentstyle[twoside,fleqn,espcrc2,epsf]{article}

\newcommand{\AmS}{{\protect\the\textfont2
  A\kern-.1667em\lower.5ex\hbox{M}\kern-.125emS}}

\def\Preprint{\vspace*{-7.5cm} 
 \noindent FTUV/96-87 \\ 
  IFIC/96-96 \\  
  \vspace{5.3cm}}
\def\refjl#1#2#3#4#5#6{\bibitem{#1} #2, {#3} {#4} (#5) #6.}
\def\refbk#1#2#3#4{\bibitem{#1} #2, {\it #3}, #4.}
\def\etal{{et al}}
\def\NP{Nucl. Phys.}
\def\NPPS{Nucl. Phys. B (Proc. Suppl.)}
\def\PL{Phys. Lett.}
\def\PRL{Phys. Rev. Lett.}
\def\PR{Phys. Rev.}
\def\PRep{Phys. Rep.}
\def\ZP{Z. Phys.}

\def\JPG{J. Phys. G: Nucl. Phys.}           

\def\APNY{Ann. Phys., NY}

\def\RPP{Rep. Prog. Phys.}

\def\PPNP{Prog. Part. Nucl. Phys.}
\def\CPC{Comput. Phys. Commun.}
\newcommand{\eqn}[1]{(\ref{#1})}
\newcommand{\be}{\begin{equation}}
\newcommand{\ee}{\end{equation}}
\newcommand{\no}{\nonumber}
\newcommand{\bel}[1]{\be\label{#1}}
\newcommand{\ba}{\begin{array}{c}}
\newcommand{\bat}{\begin{array}{cc}}
\newcommand{\ea}{\end{array}}
\newcommand{\beqn}{\begin{eqnarray}}
\newcommand{\eeqn}{\end{eqnarray}}

\newcommand{\bi}{\begin{itemize}}
\newcommand{\ei}{\end{itemize}}

\newcommand{\rms}{\rm\scriptsize}

\newcommand{\cO}{{\cal O}}
\newcommand{\cP}{{\cal P}}

\newcommand{\cA}{{\cal A}}
\newcommand{\cI}{{\cal I}}

\title{Tau Lepton Physics: Theory Overview\thanks{
        Invited talk at the Fourth International Workshop
        on Tau Lepton Physics (TAU96), Colorado, September 1996}}

\author{A. Pich    \\ \noindent  
         Departament de F\'{\i}sica Te\`orica, 
         IFIC,  CSIC --- Universitat de Val\`encia, \\ 
         Dr. Moliner 50, E--46100 Burjassot, Val\`encia, Spain}
       
\begin{document}

\begin{abstract}
The pure leptonic or semileptonic character of $\tau$
decays makes  them a good laboratory to test the structure of the
weak currents  and the universality of their couplings to the gauge
bosons. The hadronic $\tau$ decay modes constitute an
ideal  tool for  studying low--energy effects of the strong
interactions in  very clean conditions; a well--known example is the
precise determination of the QCD coupling from $\tau$--decay data. 
New physics phenomena, such as a non-zero $m_{\nu_\tau}$ or 
violations of (flavour / CP) conservation laws
can also be searched for with $\tau$ decays.
\end{abstract}


\maketitle
\Preprint
\section{INTRODUCTION}
\label{sec:introduction}

The $\tau$ lepton is a member of
the third generation which decays into particles belonging to the first
and second ones.
Thus, $\tau$ physics could provide some
clues to the puzzle of the recurring families of leptons and quarks.
In fact, one na\"{\i}vely expects the heavier fermions to be more sensitive to
whatever dynamics is responsible for the fermion--mass generation.

The pure leptonic or semileptonic character of $\tau$  decays
provides a clean laboratory to test the structure of the weak
currents  and the universality of their couplings to the gauge bosons.
Moreover, the  $\tau$ is
the only known lepton massive enough to  decay  into  hadrons;
its  semileptonic decays are then an ideal tool for studying
strong interaction effects in  very clean conditions.

 The last few years have witnessed a substantial change on our knowledge
of the $\tau$ properties \cite{montreux}.
The large (and clean) data samples collected by the most recent experiments
have improved considerably the statistical accuracy and, moreover,
have brought a new level of systematic understanding.
All experimental results obtained so far confirm the Standard Model (SM)
scenario, in which the $\tau$ is a sequential lepton with its own quantum
number and associated neutrino.

With the increased sensitivities achieved recently, 
interesting limits on possible 
new physics contributions to the $\tau$ decay amplitudes
start to emerge.
The present tests on lepton universality
will be reviewed in section \ref{sec:universality},
both for the charged and neutral current sectors. 
The Lorentz structure of the leptonic charged currents
will be discussed in section \ref{sec:lorentz}. 
The quality of the hadronic $\tau$--decay data allows to study
important properties of low--energy QCD,
involving both perturbative and 
non-perturbative aspects; this will be addressed in section~\ref{sec:QCD}.
Section~\ref{sec:new-physics} contains a
brief overview of several searches for new physics
phenomena, using $\tau$ decays. A few summarizing comments 
will be finally given in section~\ref{sec:summary}.

\section{UNIVERSALITY}      
\label{sec:universality}

\subsection{Charged Currents}
\label{subsec:cc}

The leptonic decays 
$\tau^-\to e^-\bar\nu_e\nu_\tau,\mu^-\bar\nu_\mu\nu_\tau$
are theoretically understood at the level of the electroweak
radiative corrections \cite{MS:88}.
Within the SM,
\begin{equation}
\label{eq:leptonic}
\Gamma (\tau^- \rightarrow \nu_{\tau} l^- \bar{\nu}_l)  \, = \,
  {G_F^2 m_{\tau}^5 \over 192 \pi^3} \, f(m_l^2 / m_{\tau}^2) \, 
r_{EW},
\end{equation}
where 
$f(x) = 1 - 8 x + 8 x^3 - x^4 - 12 x^2 \log{x}$.
The factor $r_{EW}=0.9960$ takes into account radiative corrections 
not included in the
Fermi coupling constant $G_F$, and the non-local structure of the
$W$ propagator \cite{MS:88}.

%
\begin{table}[thb]
\centering
\caption{Average values \protect\cite{PDG:96,Evans,Weber,Rolandi}
of some basic $\tau$ parameters.
$h^-$ stands for either $\pi^-$ or $K^-$.}
\label{tab:parameters}
\vspace{0.2cm}
\begin{tabular}{lc}
\hline 
$m_\tau$ & $(1777.00^{+0.30}_{-0.27})$ MeV \\
$\tau_\tau$ & $(290.21\pm 1.15)$ fs \\
$B_e$ & $(17.786\pm 0.072)\% $ \\
$B_\mu$ & $(17.317\pm 0.078)\% $ \\
Br($\tau^-\to\nu_\tau\pi^-$) & $(11.01\pm 0.11)\% $ \\
Br($\tau^-\to\nu_\tau K^-$) & $(0.692\pm 0.028)\% $ \\
Br($\tau^-\to\nu_\tau h^-$) & $(11.70\pm 0.11)\% $
\\ \hline
\end{tabular}
\end{table}
%
\begin{figure}[bth]
\centerline{\epsfxsize =7.5cm \epsfbox{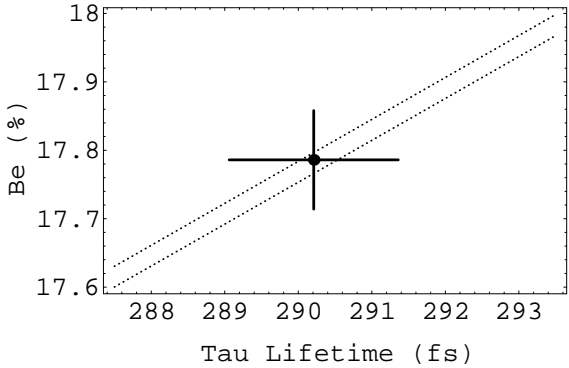}}
\vspace{-0.5cm}
\caption{Relation between $B_e$ and $\tau_\tau$. The dotted
band corresponds to the prediction in Eq.~(\protect\ref{eq:relation}).
\label{fig:BeLife}}
\end{figure}

Using the value of $G_F$  
measured in $\mu$ decay, Eq.~\eqn{eq:leptonic} 
provides a
relation \cite{PI:92} between the $\tau$ lifetime
and the leptonic branching ratios
$B_l\equiv B(\tau^-\to\nu_\tau l^-\bar\nu_l)$:
\beqn
\label{eq:relation}
B_e & = & {B_\mu \over 0.972564\pm 0.000010} 
\no\\ & = &
{ \tau_{\tau} \over (1.6321 \pm 0.0014) \times 10^{-12}\, {\rm s} } \, .
\eeqn
The errors reflect the present uncertainty of $0.3$ MeV
in the value of $m_\tau$.

The relevant experimental measurements are given in Table~\ref{tab:parameters}.
The predicted $B_\mu/B_e$ ratio is in perfect agreement with the measured
value $B_\mu/B_e = 0.974 \pm 0.006$.  As shown in
Fig.~\ref{fig:BeLife}, the relation between $B_e$ and
$\tau_\tau$ is also well satisfied by the present data. Notice,  that this
relation is very sensitive to the value of the $\tau$ mass
[$\Gamma_{\tau\to l}\propto m_\tau^5$]. The most recent measurements of
$\tau_\tau$, $B_e$ and $m_\tau$ have consistently moved the world averages
in the correct direction, eliminating the previous ($\sim 2\sigma$)
disagreement. The experimental precision (0.4\%) is already approaching the
level where a possible non-zero $\nu_\tau$ mass could become relevant; the
present bound \cite{ALEPH:95}
$m_{\nu_\tau}< 24$ MeV (95\% CL) only guarantees that such 
effect\footnote{
The preliminary ALEPH bound \protect\cite{Passalacqua},
$m_{\nu_\tau}< 18.2$ MeV (95\% CL), implies a correction smaller than
0.08\% .}
is below 0.14\%.

The decay modes $\tau^-\to\nu_\tau P^-$ [$P=\pi,K$]
can also be accurately predicted through the ratios
$R_{\tau/P}\equiv\Gamma(\tau^-\to\nu_\tau P^-)/
\Gamma(P^-\to \mu^-\bar\nu_\mu)$, 
where the dependence on the hadronic matrix elements (the so--called
decay constants $f_P$) factors out:
\bel{eq:R_tp}
R_{\tau/P}  =
{m_\tau^3\over 2 m_P m_\mu^2}
{(1-m_P^2/ m_\tau^2)^2\over
 (1-m_\mu^2/ m_P^2)^2} 
\left( 1 + \delta R_{\tau/P}\right) .
\ee
Owing to the different energy scales involved, the radiative
corrections to the $\tau^-\to\nu_\tau P^-$ amplitudes
are however not the same than the corresponding effects in
$P^-\to\mu^-\bar\nu_\mu$. The relative correction
has been estimated \cite{MS:93,DF:94} to be:
\beqn\label{eq:dR_tp_tk}
\delta R_{\tau/\pi} &=& (0.16\pm 0.14)\% \ , \no \\
\delta R_{\tau/K} &=& (0.90\pm 0.22)\%  \ .
\eeqn

All these measurements can be used to test the universality of
the $W$ couplings to the leptonic charged currents.
The $B_\mu/B_e$ ratio constraints $|g_\mu/g_e|$, while
$B_e/\tau_\tau$ and $R_{\tau/P}$
provide information on $|g_\tau/g_\mu|$.
The present results are shown in Tables \ref{tab:univme} and
\ref{tab:univtm}, together with the values obtained from the
ratio \cite{BR:92}   
$R_{\pi\to e/\mu}\equiv\Gamma(\pi^-\to e^-\bar\nu_e)/
\Gamma(\pi^-\to\mu^-\bar\nu_\mu)$,
and from the comparison of the $\sigma\cdot B$ partial production
cross-sections for the various $W^-\to l^-\bar\nu_l$ decay
modes at the $p$-$\bar p$ colliders \cite{UA1:89}. 

\begin{table}[bth]
\centering
\caption{Present constraints on $|g_\mu/g_e|$.}
\label{tab:univme}
\vspace{0.2cm}
\begin{tabular}{lc}
\hline
& $|g_\mu/g_e|$ \\ \hline
$B_\mu/B_e$ & $1.0005\pm 0.0030$
\\
$R_{\pi\to e/\mu}$ & $1.0017\pm 0.0015$
\\
$\sigma\cdot B_{W\to\mu/e}$ & $1.01\pm 0.04$
\\ \hline
\end{tabular}\vspace{1cm}
%
\caption{Present constraints on $|g_\tau/g_\mu|$.}
\label{tab:univtm}
\vspace{0.2cm}
\begin{tabular}{lc}
\hline
& $|g_\tau/g_\mu|$  \\ \hline
$B_e\tau_\mu/\tau_\tau$ & $1.0001\pm 0.0029$
\\
$R_{\tau/\pi}$ &  $1.005\pm 0.005$
\\
$R_{\tau/K}$ & $0.984\pm 0.020$
\\
$R_{\tau/h}$ & $1.004\pm 0.005$
\\
$\sigma\cdot B_{W\to\tau/\mu}$ & $0.99\pm 0.05$
\\ \hline
\end{tabular}
\end{table}
%

The present data verifies the universality of the leptonic
charged--current couplings to the 0.15\% ($e/\mu$) and 0.30\%
($\tau/\mu$) level. The precision of the most recent
$\tau$--decay measurements is becoming competitive with the 
more accurate $\pi$--decay determination. 
It is important to realize the complementarity of the
different universality tests. 
The pure leptonic decay modes probe
the charged--current couplings of a transverse $W$. In contrast,
the decays $\pi/K\to l\bar\nu$ and $\tau\to\nu_\tau\pi/K$ are only
sensitive to the spin--0 piece of the charged current; thus,
they could unveil the presence of possible scalar--exchange
contributions with Yukawa--like couplings proportional to some
power of the charged--lepton mass.
One can easily imagine new physics scenarios which would modify 
differently the two types of leptonic couplings \cite{MA:94}. 
For instance,
in the usual two Higgs doublet model, charged--scalar exchange
generates a correction to the ratio $B_\mu/B_e$, but 
$R_{\pi\to e/\mu}$ remains unaffected.
Similarly, lepton mixing between the $\nu_\tau$ and an hypothetical
heavy neutrino would not modify the ratios  $B_\mu/B_e$ and
$R_{\pi\to e/\mu}$, but would certainly correct the relation between
$B_l$ and the $\tau$ lifetime.
 
\subsection{Neutral Currents}
\label{subsec:nc}

In the SM, all leptons with equal electric charge have identical
couplings to the $Z$ boson:
$v_l = T_3^l (1-4|Q_l|\sin^2{\theta_W})$, $a_l=T_3^l$.
This has been tested at LEP and SLC \cite{Roney,LEP:96},
where the {\it effective} vector and axial--vector couplings of the three
charged leptons have been determined, 
by measuring the total $e^+e^-\to Z \to l^+l^-$
cross--section, the forward--backward asymmetry,
the (final) polarization asymmetry, the forward--backward (final) polarization
asymmetry, and (at SLC) the left--right asymmetry between the
cross--sections for initial left-- and right--handed electrons:
\goodbreak
\beqn\label{eq:LEP_obs}
\lefteqn{
\sigma^{0,l} = 
 {12 \pi  \over M_Z^2 } \, {\Gamma_e \Gamma_l\over\Gamma_Z^2}\, ,
\qquad\;
\cA_{\mbox{\rms FB}}^{0,l} = {3 \over 4} \cP_e \cP_l \, ,}
\no\\ 
\lefteqn{
\cA_{\mbox{\rms Pol}}^{0,l} = \cP_l \, ,
\qquad\qquad\quad
\cA_{\mbox{\rms FB,Pol}}^{0,l}  =  {3 \over 4} \cP_e  \, ,}
\\
\lefteqn{
\cA_{\mbox{\rms LR}}^0 =  - \cP_e \,  ,    } \no
\eeqn
where
\bel{eq:P_l}
\cP_l \, \equiv \, { - 2 v_l a_l \over v_l^2 + a_l^2} 
\ee
is the average longitudinal polarization of the lepton $l^-$.

The $Z$ partial decay width to the $l^+l^-$ final state,
\bel{eq:Z_l_QED}
\Gamma_l  = 
{G_F M_Z^3\over 6\pi\sqrt{2}} \, (v_l^2 + a_l^2)\, 
\left(1 + {3\alpha\over 4\pi}\right) ,
\ee
determines the sum $(v_l^2 + a_l^2)$, while the ratio $v_l/a_l$
is derived from the asymmetries.
The signs of $v_l$ and $a_l$ are fixed by requiring $a_e<0$.

The measurement of the final polarization asymmetries can (only) be done for 
$l=\tau$, because the spin polarization of the $\tau$'s
is reflected in the distorted distribution of their decay products.
Therefore, $\cP_\tau$ and $\cP_e$ can be determined from a
measurement of the spectrum of the final charged particles in the
decay of one $\tau$, or by studying the correlated distributions
between the final products of both $\tau's$ \cite{ABGPR:92}.

\begin{table*}[tbh]
\caption{Measured values \protect\cite{LEP:96}
of $\Gamma_l\equiv\Gamma(Z\to l^+l^-)$
and the leptonic forward--backward asymmetries.
The last column shows the combined result 
(for a massless lepton) assuming lepton universality.
\label{tab:LEP_asym}}
\vspace{0.2cm}
\begin{tabular*}{\textwidth}{@{}l@{\extracolsep{\fill}}cccc}
\hline
& $e$ & $\mu$ & $\tau$ & $l$ 
\\ \hline
$\Gamma_l$ \, (MeV) & $83.96\pm 0.15$
& $83.79\pm 0.22$ & $83.72\pm 0.26$ & $83.91\pm 0.11$
\\
$\cA_{\mbox{\rms FB}}^{0,l}$ \, (\%) & $1.60\pm 0.24$
& $1.62\pm 0.13$ & $2.01\pm 0.18$ & $1.74\pm 0.10$
\\ \hline \\
\end{tabular*}
%
\caption{Measured values \protect\cite{LEP:96}
of the different polarization asymmetries.}
\label{tab:pol_asym}
\vspace{0.2cm}
\begin{tabular*}{\textwidth}{@{}c@{\extracolsep{\fill}}ccc}
\hline
$\cA_{\mbox{\rms Pol}}^{0,\tau} = \cP_\tau$ &
${4\over 3}\cA^{0,\tau}_{\mbox{\rms FB,Pol}} = \cP_e$ &
$-\cA_{\mbox{\rms LR}}^0 = \cP_e$
& $- \{{4\over 3}\cA_{\mbox{\rms FB}}^{0,l}\}^{1/2} = P_l$
\\ \hline
$-0.1401\pm 0.0067$ & $-0.1382\pm 0.0076$ & $-0.1542\pm 0.0037$
& $-0.1523\pm 0.0044$
\\ \hline
\end{tabular*}
\end{table*}

Tables~\ref{tab:LEP_asym} and \ref{tab:pol_asym}
show the present experimental results
for the leptonic $Z$--decay widths and asymmetries.
The data are in excellent agreement with the SM predictions
and confirm the universality of the leptonic neutral couplings\footnote{
A small 0.2\% difference between $\Gamma_\tau$ and $\Gamma_{e,\mu}$
is generated by the $m_\tau$ corrections.}.
There is however a small ($\sim 2\sigma$) discrepancy between the
$\cP_e$ values obtained \cite{LEP:96} from 
$\cA^{0,\tau}_{\mbox{\rms FB,Pol}}$ 
and $\cA_{\mbox{\rms LR}}^0$.
Assuming lepton universality, 
the combined result from all leptonic asymmetries gives
\bel{eq:average_P_l}
\cP_l = - 0.1500\pm 0.0025 \ .
\ee

The measurement of $\cA_{\mbox{\rms Pol}}^{0,\tau}$ and
$\cA^{0,\tau}_{\mbox{\rms FB,Pol}}$ assumes that the $\tau$ decay
proceeds through the SM charged--current interaction.
A more general analysis should take into account the fact that the
$\tau$--decay width depends on the product $\xi\cP_\tau$ 
(see section~\ref{sec:lorentz}), 
where $\xi$
is the corresponding Michel parameter in leptonic decays, or
the equivalent quantity $\xi_h$  ($=h_{\nu_\tau}$) in the semileptonic 
modes.
A separate measurement of $\xi$ and $\cP_\tau$ has been performed by
ALEPH \cite{ALEPH:94} ($\cP_\tau = -0.139\pm 0.040$)
and L3 \cite{L3:96} ($\cP_\tau = -0.154\pm 0.022$),
using the correlated distribution of the $\tau^+\tau^-$ decays.

\begin{figure}[tbh]
\centerline{\epsfxsize =7.5cm \epsfbox{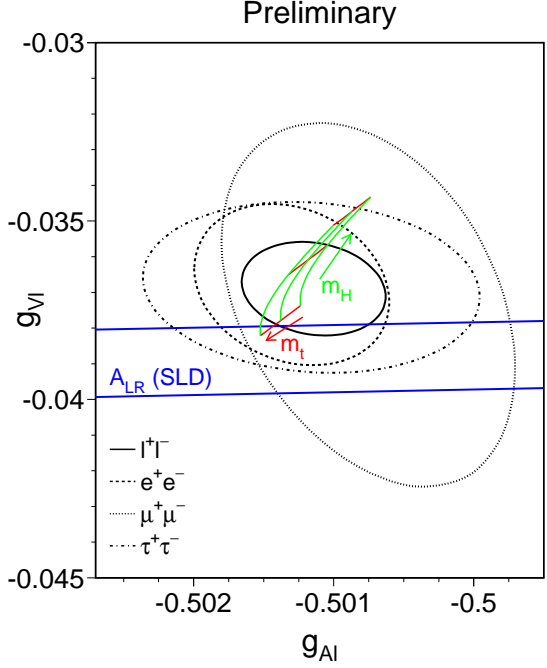}}
\vspace{-0.5cm}
\caption{68\% probability contours in the $a_l$-$v_l$ plane
from LEP measurements \protect\cite{LEP:96}. 
The solid contour assumes lepton universality. 
Also shown is the $1\sigma$ band resulting from the
$\protect\cA_{\mbox{\protect\rms LR}}^0$ measurement at SLD. 
The grid corresponds to the SM prediction.} 
\label{fig:gagv}
\end{figure}

The combined analysis of all leptonic observables from LEP
and SLD ($\cA_{\mbox{\rms LR}}^0$) results in the
effective vector and axial--vector couplings given in
Table~\ref{tab:nc_measured} \cite{LEP:96}. 
The corresponding 68\% probability contours in the $a_l$--$v_l$ plane 
are shown in Fig.~\ref{fig:gagv}.
The measured ratios of the $e$, $\mu$ and $\tau$ couplings
provide a test of charged--lepton universality in the neutral--current 
sector.

The neutrino coupling can be determined from the invisible 
$Z$--decay width, by assuming three identical neutrino generations
with left--handed couplings (i.e., $v_\nu=a_\nu$), 
and fixing the sign from neutrino scattering 
data \cite{CHARMII:94}.
The resulting experimental value \cite{LEP:96},
given in Table~\ref{tab:nc_measured},
is in perfect agreement with the SM.
Alternatively, one can use the SM prediction for 
$\Gamma_{\mbox{\rms inv}}/\Gamma_l$
to get a determination of the number of (light) neutrino flavours
\cite{LEP:96}:
$N_\nu = 2.989\pm 0.012$.
The universality of the neutrino couplings has been tested
with $\nu_\mu e$ scattering data, which fixes \cite{CHARMII:94b}
the $\nu_\mu$ coupling to the $Z$: \ 
$v_{\nu_\mu} =  a_{\nu_\mu} = 0.502\pm 0.017$.

The measured leptonic asymmetries can be used to obtain the
effective electroweak mixing angle in the charged--lepton sector: 
\cite{LEP:96}
\bel{eq:bar_s_W_l}
\sin^2{\theta^{\mbox{\rms lept}}_{\mbox{\rms eff}}} \equiv
{1\over 4}   \left( 1 - {v_l\over a_l}\right)  = 0.23114\pm 0.00031 \, .
\ee
Including also the hadronic asymmetries, one gets \cite{LEP:96}
$\sin^2{\theta^{\mbox{\rms lept}}_{\mbox{\rms eff}}} =
0.23165\pm 0.00024$ 
with a $\chi^2/\mbox{\rm d.o.f.} = 12.8/6$.

\begin{table}[bht]
\centering
\caption{
Effective vector and axial--vector lepton couplings
derived from LEP and SLD data \protect\cite{LEP:96}.
\label{tab:nc_measured}}
\vspace{0.2cm}
\begin{tabular}{ccc}     
\hline
$v_e$ & 
        $-0.03828 \pm 0.00079$
\\
$v_\mu$ & 
          $-0.0358 \pm 0.0030$
\\
$v_\tau$ & 
           $-0.0367 \pm 0.0016$
\\
$a_e$ & 
        $-0.50119 \pm 0.00045$
\\
$a_\mu$ & 
          $-0.50086 \pm 0.00068$
\\
$a_\tau$ & 
           $-0.50117 \pm 0.00079$
\\ \hline  
$v_\mu/v_e$ & 
              $\phantom{-}0.935\pm 0.085$
\\
$v_\tau/v_e$ & 
               $\phantom{-}0.959\pm 0.046$
\\
$a_\mu/a_e$ & 
              $\phantom{-}0.9993\pm 0.0017$
\\
$a_\tau/a_e$ & 
               $\phantom{-}1.0000\pm 0.0019$
\\ \hline
\multicolumn{2}{c}{With Lepton Universality}\\ \hline
$v_l$ & 
        $-0.03776 \pm 0.00062$
\\
$a_l$ & 
        $-0.50108 \pm 0.00034$
\\
$a_\nu=v_\nu$ & 
                $+0.5009\pm 0.0010$ 
\\ \hline
\end{tabular}
\end{table}

\section{LORENTZ STRUCTURE}  
\label{sec:lorentz}

Let us consider the 
decay $l^-\to\nu_l l'^-\bar\nu_{l'}$, 
where the lepton pair ($l$, $l^\prime $)
may be ($\mu$, $e$), ($\tau$, $e$), or ($\tau$, $\mu$). 
The most general, local, derivative--free, lepton--number conserving, 
four--lepton interaction Hamiltonian, 
consistent with locality and Lorentz invariance
\cite{MI:50,BM:57,SCH:83,FGJ:86,FG:93,PS:95},
\be
{\cal H} = 4 \frac{G_{l'l}}{\sqrt{2}}
\sum_{n,\epsilon,\omega}          
g^n_{\epsilon\omega}   
\left[ \overline{l'_\epsilon} 
\Gamma^n {(\nu_{l'})}_\sigma \right]\, 
\left[ \overline{({\nu_l})_\lambda} \Gamma_n 
	l_\omega \right]\ ,
\label{eq:hamiltonian}
\ee
contains ten complex coupling constants or, since a common phase is
arbitrary, nineteen independent real parameters
which could be different for each leptonic decay.
The subindices
$\epsilon , \omega , \sigma, \lambda$ label the chiralities (left--handed,
right--handed)  of the  corresponding  fermions, and $n$ the
type of interaction: 
scalar ($I$), vector ($\gamma^\mu$), tensor 
($\sigma^{\mu\nu}/\sqrt{2}$).
For given $n, \epsilon ,
\omega $, the neutrino chiralities $\sigma $ and $\lambda$
are uniquely determined.

Taking out a common factor $G_{l'l}$, which is determined by the total
decay rate, the coupling constants $g^n_{\epsilon\omega}$
are normalized to \cite{FGJ:86}
\beqn\label{eq:normalization}
1 &\!\!\! = &\!\!\!
{1\over 4} \,\left( |g^S_{RR}|^2 + |g^S_{RL}|^2
    + |g^S_{LR}|^2 + |g^S_{LL}|^2 \right)
\no \\ & &\!\!\! \mbox{}
+ \left(
   |g^V_{RR}|^2 + |g^V_{RL}|^2 + |g^V_{LR}|^2 + |g^V_{LL}|^2 \right)
\no\\ & &\!\!\! \mbox{}
   +  3 \,\left( |g^T_{RL}|^2 + |g^T_{LR}|^2 \right) 
\, .
\eeqn
In the SM, $g^V_{LL}  = 1$  and all other
$g^n_{\epsilon\omega} = 0 $.

For an initial lepton polarization ${\cal P}_l$,
the final charged--lepton distribution in the decaying--lepton 
rest frame
is usually parameterized \cite{BM:57} in the form  
\beqn\label{eq:spectrum}
\lefteqn{{d^2\Gamma \over dx\, d\cos\theta} =
{m_l\omega^4 \over 2\pi^3} G_{l'l}^2 \sqrt{x^2-x_0^2}}
\no\\ &&
\times\left\{ F(x) - 
{\xi\over 3}\, {\cal P}_l\,\sqrt{x^2-x_0^2}
\,\cos{\theta}\, A(x)\right\} ,
\eeqn
where $\theta$ is the angle between the $l^-$ spin and the
final charged--lepton momentum,
$\, \omega \equiv (m_l^2 + m_{l'}^2)/2 m_l \, $
is the maximum $l'^-$ energy for massless neutrinos, $x \equiv E_{l'^-} /
\omega$ is the reduced energy, $x_0\equiv m_{l'}/\omega$
and
\beqn\label{eq:Fx_Ax_def}
F(x)  &\!\!\! = &\!\!\! 
  x (1 - x) + {2\over 9} \rho 
 \left(4 x^2 - 3 x - x_0^2 \right) 
\no\\ &&\!\!\!\!  \mbox{}
+  \eta\, x_0 (1-x)
\, , \\
A(x) &\!\!\! = &\!\!\! 
 1 - x 
  + {2\over 3}  \delta \left( 4 x - 4 + \sqrt{1-x_0^2}  
\right)  \, .\no
\eeqn

For unpolarized $l's$, the distribution is characterized by
the so-called Michel \cite{MI:50} parameter $\rho$
and the low--energy parameter $\eta$. Two more parameters, $\xi$
and $\delta$, can be determined when the initial lepton polarization is known.
If the polarization of the final charged lepton is also measured,
5 additional independent parameters \cite{PDG:96}  
($\xi'$, $\xi''$, $\eta''$, $\alpha'$, $\beta'$)
appear. 

For massless neutrinos,
the total decay rate is still given
by Eq.~\eqn{eq:leptonic},
but changing $G_F$ to \cite{PS:95}
\be\label{eq:gamma}
\widehat{G}_{l'l}  \equiv  G_{l'l} \,
\sqrt{1 + 4\,\eta\, {m_{l'}\over m_l}\,
{g\!\left( m_{l'}^2/ m_l^2 \right)\over  
f\!\left( m_{l'}^2/ m_l^2 \right)}}
\, ,
\ee
where
$g(z) = 1 + 9 z - 9 z^2 - z^3 + 6 z (1+z) \ln{z}$.
Thus, $\widehat{G}_{e\mu}$ corresponds to the Fermi coupling 
$G_F$, measured in $\mu$ decay.
The $B_\mu/B_e$ and $B_e\tau_\mu/\tau_\tau$
universality tests, discussed in the previous section,
actually prove the ratios
$|\widehat{G}_{\mu\tau}/\widehat{G}_{e\tau}|$
and $|\widehat{G}_{e\tau}/\widehat{G}_{e\mu}|$, respectively.
An important point, emphatically stressed by
Fetscher and Gerber \cite{FG:93}, concerns the extraction
of $G_{e \mu}$, whose uncertainty is dominated
by the uncertainty in $\eta_{\mu\to e}$. 

In terms of the $g_{\epsilon\omega}^n$
couplings, the shape parameters in Eqs.~\eqn{eq:spectrum}
and \eqn{eq:Fx_Ax_def}
are:
\beqn\label{eq:michel}
\lefteqn{\rho = {3\over 4} (\beta^+ + \beta^-) + (\gamma^+ + \gamma^-) \, ,}
\no\\
\lefteqn{\xi = 3 (\alpha^- - \alpha^+) + (\beta^- - \beta^+)
  + {7\over 3} (\gamma^+ - \gamma^-) \, ,}
\no\\
\lefteqn{\xi\delta = {3\over 4} (\beta^- - \beta^+) + (\gamma^+ - \gamma^-) 
\, ,}
\\
\lefteqn{\eta =
\frac{1}{2} \mbox{\rm Re}\left[
g^V_{LL} g^{S\ast}_{RR} + g^V_{RR}  g^{S\ast}_{LL}
+ g^V_{LR} \left(g^{S\ast}_{RL} + 6 g^{T\ast}_{RL}\right)
\right. }\no\\ &&\left.\qquad\mbox{} +
g^V_{RL} \left(g^{S\ast}_{LR} + 6 g^{T\ast}_{LR}\right) 
\right] , \no
\eeqn
where \cite{Rouge}
\beqn\label{eq:abg_def}
\lefteqn{
\alpha^+ \equiv {|g^V_{RL}|}^2 + {1\over 16} {|g^S_{RL} + 6 g^T_{RL}|}^2
\, , }\no\\
\lefteqn{\beta^+\equiv {|g^V_{RR}|}^2 + {1\over 4} {|g^S_{RR}|}^2
\, , }\\
\lefteqn{\gamma^+\equiv {3\over 16} {|g^S_{RL} - 2 g^T_{RL}|}^2
\, , }\no
\eeqn
are positive--definite combinations of decay constants, corresponding to 
a final right--handed lepton, while 
$\alpha^-$, $\beta^-$, $\gamma^-$ denote the corresponding combinations
with opposite chiralities ($R\leftrightarrow L$).
In the SM, $\rho = \delta = 3/4$, 
$\eta = \eta'' = \alpha' = \beta' = 0 $ and 
$\xi = \xi' = \xi'' = 1 $.

The normalization constraint \eqn{eq:normalization} is equivalent to
$\alpha^+ + \alpha^- + \beta^+ + \beta^- + \gamma^+ + \gamma^- = 1$.
It is convenient to introduce \cite{FGJ:86} the probabilities
$Q_{\epsilon\omega}$ for the
decay of an $\omega$--handed $l^-$
into an $\epsilon$--handed 
daughter lepton,
\beqn\label{eq:Q_LL}
\lefteqn{Q_{LL}  = \beta^- = 
{1 \over 4} |g^S_{LL}|^2 \! +  |g^V_{LL}|^2 
\, , }\no\\ 
\lefteqn{Q_{RR} = \beta^+ =
{1 \over 4} |g^S_{RR}|^2 \! + \! |g^V_{RR}|^2 
\, , }\\ 
\lefteqn{Q_{LR} = \alpha^- + \gamma^- =
{1 \over 4} |g^S_{LR}|^2 \! + \!  |g^V_{LR}|^2
            \! + \!   3 |g^T_{LR}|^2 
\, , }\no\\ 
\lefteqn{Q_{RL} = \alpha^+ + \gamma^+ =
{1 \over 4} |g^S_{RL}|^2  \! + \!  |g^V_{RL}|^2
            \! + \!  3 |g^T_{RL}|^2
\, . }\no  
\end{eqnarray}
Upper bounds on any of these (positive--semidefinite) probabilities 
translate into corresponding limits for all couplings with the 
given chiralities.

\begin{figure}[tbh]
\centerline{\epsfxsize =7.5cm \epsfbox{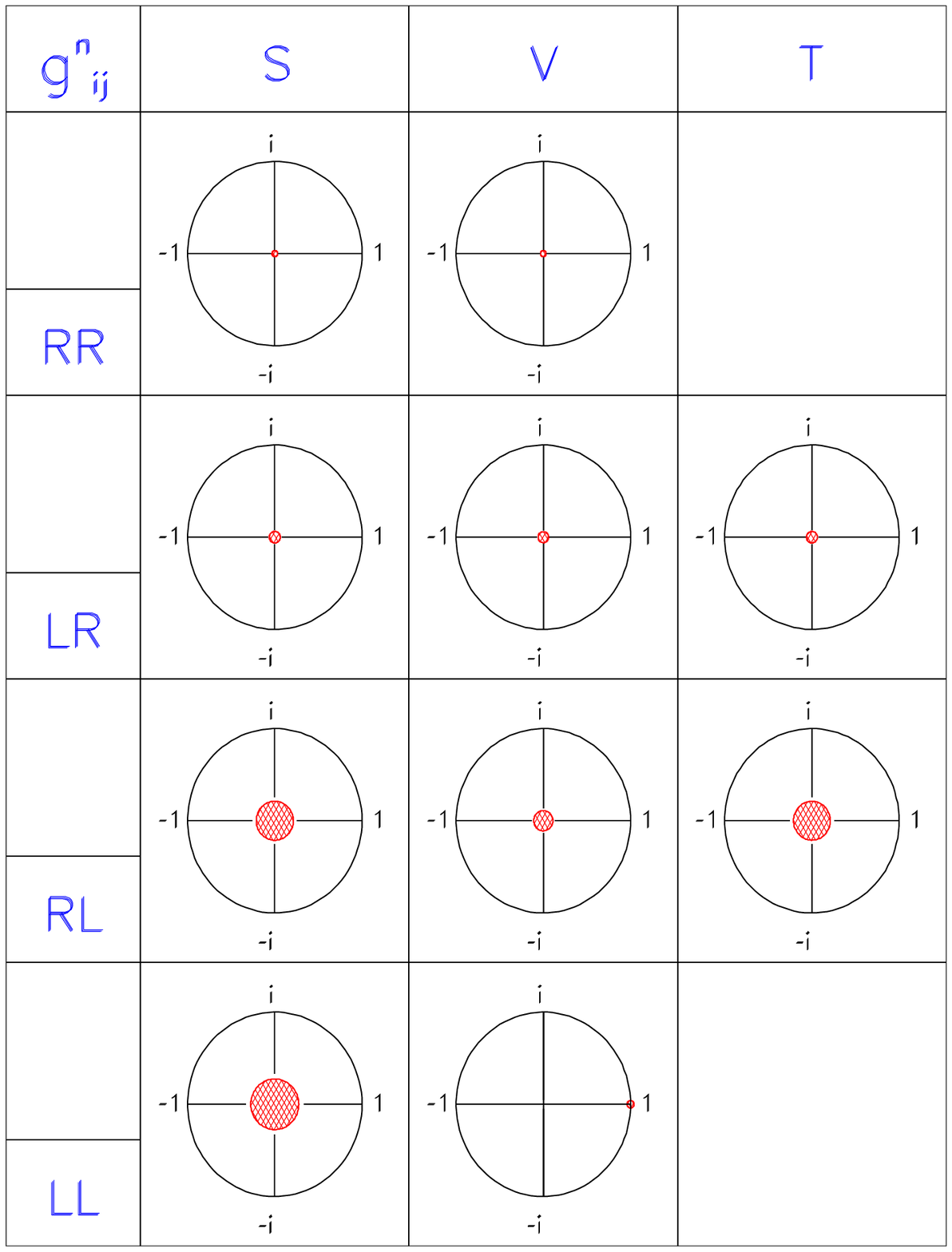}}
\vspace{-1.5cm}
\caption{90\% CL experimental limits  \protect\cite{PDG:96} 
for the normalized $\mu$--decay couplings
$g'^n_{\epsilon\omega }\equiv g^n_{\epsilon\omega }/ N^n$,
where
$N^n \equiv \protect\mbox{\rm max}(|g^n_{\epsilon\omega }|) =2$,
1, $1/\protect\sqrt{3} $ for $n =$ S, V, T.
(Taken from Ref.~\protect\cite{LR:95}).}
\label{fig:mu_couplings}
\end{figure}

For $\mu$ decay, where precise measurements of the polarizations of
both $\mu$ and $e$ have been performed, there exist \cite{FGJ:86}
upper bounds on $Q_{RR}$, $Q_{LR}$ and $Q_{RL}$, and a lower bound
on $Q_{LL}$. They imply corresponding upper bounds on the 8
couplings $|g^n_{RR}|$, $|g^n_{LR}|$ and $|g^n_{RL}|$.
The measurements of the $\mu^-$ and the $e^-$ do not allow to
determine $|g^S_{LL}|$ and $|g^V_{LL}|$ separately \cite{FGJ:86,JA:66}.
Nevertheless, since the helicity of the $\nu_\mu$ in pion decay is
experimentally known \cite{FE:84}    
to be $-1$, a lower limit on $|g^V_{LL}|$ is
obtained \cite{FGJ:86} from the inverse muon decay
$\nu_\mu e^-\to\mu^-\nu_e$.
The present (90\% CL) bounds \cite{PDG:96}     
on the $\mu$--decay couplings
are shown in Fig.~\ref{fig:mu_couplings}. 
These limits show nicely 
that the bulk of the $\mu$--decay transition amplitude is indeed of
the predicted V$-$A type.

\begin{figure}[tbh]
\centerline{\epsfxsize =7.5cm \epsfbox{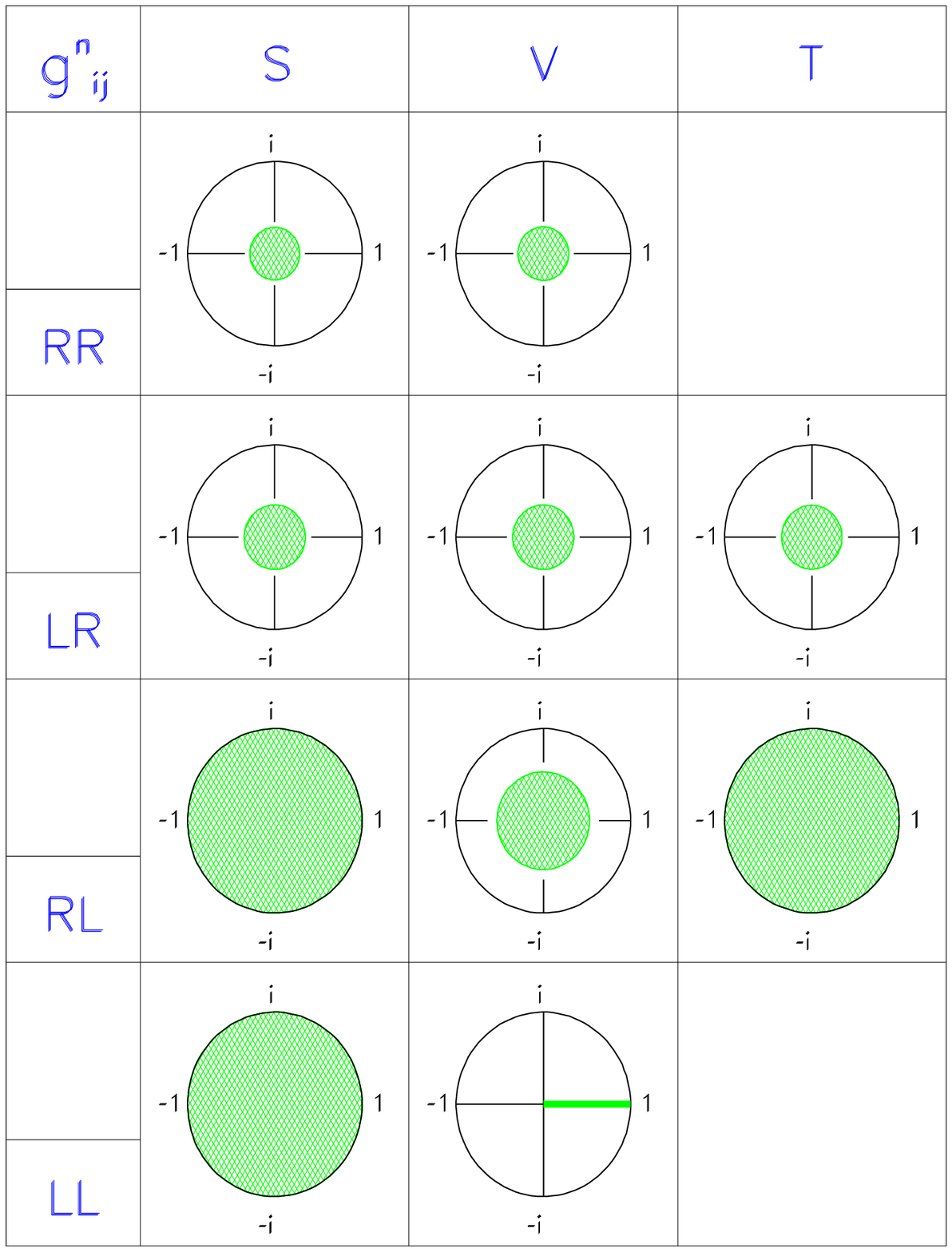}}
\vspace{-1.5cm}
\caption{90\% CL experimental limits
for the normalized $\tau$--decay couplings
$g'^n_{\epsilon\omega }\equiv g^n_{\epsilon\omega }/ N^n$,
assuming $e/\mu$ universality.
\label{fig:tau_couplings}}
\end{figure}

The experimental analysis of the $\tau$--decay parameters is 
necessarily
different from the one applied to the muon, because of the much
shorter $\tau$ lifetime.
The measurement of the $\tau$ polarization and the parameters
$\xi$ and $\delta$ 
is still possible due to the fact that the spins
of the $\tau^+\tau^-$ pair produced in $e^+e^-$ annihilation 
are strongly correlated
\cite{ABGPR:92,TS:71,PS:77,NE:91,FE:90,BPR:91,DDDR:93}.
Another possibility is to use
the beam polarization, as done by SLD \cite{Quigley}.
However,
the polarization of the charged lepton emitted in the $\tau$ decay
has never been measured. In principle, this could be done
for the decay $\tau^-\to\mu^-\bar\nu_\mu\nu_\tau$ by stopping the
muons and detecting their decay products \cite{FE:90,SV:96}.
The measurement of the inverse decay $\nu_\tau l^-\to\tau^-\nu_l$
looks far out of reach.

The present experimental status \cite{Evans}
on the $\tau$--decay Michel parameters
is shown in Table~\ref{tab:tau_michel}.
For comparison, the values measured in $\mu$ decay \cite{PDG:96}
are also given.
The improved accuracy of the most recent experimental analyses
has brought an enhanced sensitivity to the different shape parameters,
allowing the first measurements of $\eta_{\tau\to\mu}$, 
$\xi_{\tau\to e}$, $\xi_{\tau\to\mu}$, $(\xi\delta)_{\tau\to e}$ and 
$(\xi\delta)_{\tau\to\mu}$ \cite{Evans,Quigley,ALEPH:95b,Chadha},
without any $e/\mu$ universality assumption.

\begin{table*}[bth]
\setlength{\tabcolsep}{1.5pc}
\newlength{\digitwidth} \settowidth{\digitwidth}{\rm 0}
\catcode`?=\active \def?{\kern\digitwidth}
\caption{World average
\protect\cite{PDG:96,Evans}
Michel parameters. 
The last column ($\tau\to l$) assumes identical couplings
for $l=e,\mu$.
$\xi_{\mu\to e}$ refers to the product $\xi_{\mu\to e}\cP_\mu$,
where $\cP_\mu\approx 1$ is the longitudinal polarization
of the $\mu$ from $\pi$ decay.}
\label{tab:tau_michel}
\vspace{0.2cm}
\begin{tabular*}{\textwidth}{@{}l@{\extracolsep{\fill}}cccc}
\hline
& $\mu\to e$ & $\tau\to\mu$ & $\tau\to e$ & $\tau\to l$ 
\\ \hline
$\rho$ & $0.7518\pm 0.0026$ & $0.733\pm 0.031$ & $0.734\pm 0.016$ & 
$0.741\pm 0.014$ 
\\
$\eta$ & $-0.007\pm 0.013\phantom{-}$ & $-0.04\pm 0.20\phantom{-}$ & --- & 
$0.047\pm 0.076$
\\
$\xi$ & $1.0027\pm 0.0085$ & $1.19\pm 0.18$ & $1.09\pm 0.16$ & 
$1.04\pm 0.09$ 
\\
$\xi\delta$ & $0.7506\pm 0.0074$ & $0.73\pm 0.11$ & $ 0.80\pm 0.18$ & 
$ 0.73\pm 0.07$ 
\\ \hline \\
\end{tabular*}
\caption{90\% CL limits
for the $g^n_{\epsilon\omega}$ couplings.}
\label{table:g_tau_bounds}
\vspace{0.2cm}
\begin{tabular*}{\textwidth}{@{}l@{\extracolsep{\fill}}llll}
\hline
& \hfil $\mu\to e$\hfil &
\hfil $\tau\to\mu$\hfil &\hfil $\tau\to e$ \hfil & 
\hfil $\tau\to l$ \hfil
\\ \hline
$|g^S_{RR}|$  & $< 0.066$ & $< 0.71$ & $< 0.83$ & $< 0.57$
\\
$|g^S_{LR}|$  & $< 0.125$ & $< 0.90$ & $< 1.00$ & $< 0.70$
\\
$|g^S_{RL}|$  & $< 0.424$ & $\leq 2$ & $\leq 2$ & $\leq 2$
\\
$|g^S_{LL}|$  & $< 0.55$  & $\leq 2$ & $\leq 2$ & $\leq 2$
\\ \hline
$|g^V_{RR}|$  & $< 0.033$ & $< 0.36$ & $< 0.42$ & $< 0.29$
\\
$|g^V_{LR}|$  & $< 0.060$ & $< 0.45$ & $< 0.50$ & $< 0.35$
\\
$|g^V_{RL}|$  & $< 0.110$ & $< 0.56$ & $< 0.54$ & $< 0.53$
\\
$|g^V_{LL}|$  & $> 0.96$  & $\leq 1$ & $\leq 1$ & $\leq 1$
\\ \hline
$|g^T_{LR}|$  & $< 0.036$ & $< 0.26$ & $< 0.29$ & $< 0.20$
\\
$|g^T_{RL}|$  & $< 0.122$ & $\leq 1/\sqrt{3}$ & $\leq 1/\sqrt{3}$
              & $\leq 1/\sqrt{3}$
\\ \hline
\end{tabular*}
\end{table*}

The determination of the $\tau$ polarization parameters 
allows us to bound the total probability for the decay of
a right--handed $\tau$ \cite{FE:90},
\be\label{eq:Q_R}
Q_{\tau_R} \equiv 
Q_{RR} + Q_{LR}
= \frac{1}{2}\, \left[ 1 + \frac{\xi}{3} - \frac{16}{9} 
(\xi\delta)\right]
\; .
\ee
One finds (ignoring possible correlations among the measurements):
\begin{eqnarray}
Q_{\tau_R}^{\tau\to\mu} &\!\!\! =&\!\!\! \phantom{-}0.05\pm 0.10 \; 
< \, 0.20 \quad (90\%\;\mbox{\rm CL})\, , \no\\
Q_{\tau_R}^{\tau\to e} &\!\!\! =&\!\!\! -0.03\pm 0.16 \; 
< \, 0.25 \quad (90\%\;\mbox{\rm CL})\, , \\
Q_{\tau_R}^{\tau\to l} &\!\!\! =&\!\!\! \phantom{-}0.02\pm 0.06 \;
< \, 0.12 \quad (90\%\;\mbox{\rm CL})\, , \no
\end{eqnarray}
where the last value refers to the $\tau$ decay into either $l=e$ or $\mu$,
assuming identical $e$/$\mu$ couplings.
Since these probabilities are positive semidefinite quantities, they imply
corresponding limits on all
$|g^n_{RR}|$ and $|g^n_{LR}|$ couplings. 

A measurement of the final lepton polarization could be even more efficient,
since the total probability for the decay into a right--handed lepton
depends on a single Michel parameter:
\bel{eq:Qxi}
Q_{l'_R} \equiv Q_{RR} +  Q_{RL}
= {1\over 2} ( 1 -\xi') \, .
\ee
Thus, a single polarization measurement could bound the five RR and RL
complex couplings.

Another useful positive--definite quantity is \cite{LR:95}
\bel{eq:rxd}
\rho - \xi\delta = {3\over 2} \beta^+ + 2 \gamma^- \, ,
\ee
which provides direct bounds on $|g^V_{RR}|$ and $|g^S_{RR}|$.
A rather weak upper limit on $\gamma^+$ is obtained from the
parameter $\rho$. More stringent is the bound on $\alpha^+$ obtained from
$(1-\rho)$, which is also positive--definite; it implies a corresponding
limit on $|g^V_{RL}|$.

Table~\ref{table:g_tau_bounds} gives the resulting (90\% CL) bounds on the 
$\tau$--decay couplings.
The relevance of these limits can be better appreciated in 
Fig.~\ref{fig:tau_couplings}, where $e$/$\mu$ universality has been assumed.

If lepton universality is assumed, 
the leptonic decay ratios $B_\mu/B_e$  and $B_e\tau_\mu/\tau_\tau$
provide limits on the low--energy parameter $\eta$.  
The best sensitivity \cite{ST:94} comes from
$\widehat{G}_{\mu\tau}$,
where the term proportional to $\eta$ is not suppressed by
the small $m_e/m_l$ factor. The measured $B_\mu/B_e$ ratio implies
then:
\be\label{eq:eta_univ}
\eta_{\tau\to l} \, = \, 0.005\pm 0.027  \ .
\ee
This determination is more accurate that the one in 
Table~\ref{tab:tau_michel},
obtained from the shape of the energy distribution,
and is comparable to the value measured in $\mu$ decay.

A non-zero value of $\eta$ would show that there are at least two
different couplings with opposite chiralities for the charged leptons.
Assuming the V$-$A coupling $g_{LL}^V$ to be dominant, the
second one would be \cite{FE:90} a Higgs--type coupling $g^S_{RR}$.
To first order in new physics contributions,
$\eta\approx\mbox{\rm Re}(g^S_{RR})/2$;
Eq.~(\ref{eq:eta_univ}) puts then the (90\% CL) bound:
$-0.08 \, <\mbox{\rm Re}(g^S_{RR}) < 0.10$.

High--precision measurements of the $\tau$ decay parameters have the
potential to find signals for new phenomena. The accuracy
of the present data is still not good enough to provide
strong constraints; nevertheless, it shows
that the SM gives indeed the dominant contribution to the decay
amplitude.
Future experiments should then
look for small deviations of the SM predictions and find out the
possible source of any detected discrepancy.

\begin{table}[tbh]
\centering
\caption{Changes in the Michel parameters induced by
the addition of a single intermediate boson exchange 
($V^+$, $S^+$, $V^0$, $S^0$)
to the SM contribution \protect\cite{PS:95}}
\label{tab:summary}
\vspace{0.2cm}
\begin{tabular}{cccccc}
\hline
&  & $V^+$  &  $S^+$  &  $V^0$  &  $S^0$
\\ \hline
$\rho - 3/4$ &     & $< 0$  &  0   &  0   &  $< 0$
\\ 
$\xi - 1$    &    &   $\pm$   & $< 0$ & $< 0$ &  $\pm$ 
\\ 
$\delta\xi-3/4$& & $< 0$ & $< 0$ & $< 0$ & $< 0$
\\ 
$\eta$        &   &   0   &  $\pm$   &   $\pm$  &  $\pm$ 
\\ \hline
\end{tabular}
\end{table}

In a first analysis, it seems natural to assume \cite{PS:95}
that new physics effects would be dominated by the exchange of a single
intermediate boson, coupling to two leptonic currents.
%
Table~\ref{tab:summary}
summarizes the expected changes
on the measurable shape parameters \cite{PS:95},
in different new physics scenarios.
The four general cases studied correspond to adding a single intermediate
boson exchange, $V^+$, $S^+$, $V^0$, $S^0$ 
(charged/neutral, vector/scalar), to the SM contribution
(a non-standard $W$ would be a particular case of the
SM + $V^+$ scenario).

\section{QCD TESTS}
\label{sec:QCD}

The $\tau$ is the only presently known lepton massive enough to decay
into hadrons. Its semileptonic decays are then an ideal laboratory
for studying the hadronic weak currents in very clean conditions.
The decay modes $\tau^-\to\nu_\tau H^-$
probe the  matrix
element of the left--handed charged current between the vacuum and the
final hadronic state $H^-$,
\bel{eq:Had_matrix}
\langle H^-| \bar d_\theta \gamma^\mu (1-\gamma_5) u | 0 \rangle\, .
\ee
Contrary to the well--known process\ $e^+e^-\to\gamma\to$ hadrons,
which only tests the electromagnetic vector current, the semileptonic
$\tau$ decay modes offer the possibility to study the properties of both
vector and axial--vector currents.
 
For the decay modes with lowest multiplicity,
$\tau^-\to\nu_\tau\pi^-$  and $\tau^-\to\nu_\tau K^-$, the  relevant
matrix  elements  are  already  known  from  the  measured  decays
$\pi^-\to\mu^-\bar\nu_\mu$  and  $K^-\to\mu^-\bar\nu_\mu$.
%
%
The corresponding $\tau$ decay widths can then be predicted
rather accurately [Eq.~\eqn{eq:R_tp}].
As shown in Table~\ref{tab:univtm}, these predictions are in good 
agreement with the measured values, and provide a quite precise test
of charged--current universality.

Alternatively, the measured ratio between the
$\tau^-\to\nu_\tau K^-$ and $\tau^-\to\nu_\tau \pi^-$
decay widths can be used to obtain a value for
$\tan^2{\theta_C}\, (f_K / f_\pi)^2$:
\bel{eq:theta_C}
\left|{V_{us}\over V_{ud}}\right|^2
\, \left( {f_K \over f_\pi} \right)^2  =
(7.2 \pm 0.3) \times 10^{-2} \, . 
\ee
This number is consistent with (but less precise than) the result
$(7.67 \pm 0.06) \times 10^{-2}$  obtained from \cite{PDG:96} 
$\Gamma (K^- \to \mu^-\bar\nu_\mu ) / \Gamma (\pi^-\to\mu^-\bar\nu_\mu )$.

For the Cabibbo--allowed modes with $J^P = 1^-$, the matrix element of
the vector charged current can also be obtained, 
through an isospin  rotation, from the
isovector  part  of  the $e^+ e^-$ annihilation  cross--section  into
hadrons, which
measures the hadronic matrix element of   the  $I=1$  component
of the electromagnetic current,
\bel{eq:em_matrix}
\langle V^0|(\bar u \gamma^\mu u - \bar d \gamma^\mu d)|0\rangle\, .
\ee
The $\tau\to \nu_\tau V^-$ decay width is then expressed as an integral over
the corresponding $e^+ e^-$ cross-section \cite{TS:71,ThS:71}
$\sigma(s)\equiv\sigma^{I=1}_{e^+ e^- \to V^0}(s)$:
\beqn\label{eq:cvc}
\lefteqn{R_{\tau\to V}  \equiv
{\Gamma (\tau^-\to\nu_\tau V^-) \over \Gamma_{\tau\to e}}
=  {3 \cos^2{\theta_C} \over 2 \pi \alpha^2 m_\tau^8 }\, S_{EW}\, \cI }\, ,
&&\no\\
\lefteqn{\cI =
   \int_0^{m_\tau^2} \, ds \, (m_\tau^2 - s)^2 (m_\tau^2 + 2 s) \, s \,
 \sigma(s)\, , }
\eeqn
where the factor $S_{EW}=1.0194$ contains the renormalization--group 
improved
electroweak correction at the leading logarithm approximation \cite{MS:88}.
Using the available \ $e^+ e^- \to$ hadrons \ data,
one can then predict the $\tau$ decay widths for these modes
\cite{GRM:85,KS:90,EI:95,NP:93,Eidelman}.

\begin{table}[htb]
\centering
\caption{$R_{\tau\to V}$
from $\tau$--decay \protect\cite{PDG:96,Evans} and $e^+e^-$ data
\protect\cite{Eidelman}.}
\label{tab:vector}
\vspace{0.2cm}
\begin{tabular}{lll}
\hline
$V^-$   & \multicolumn{2}{c}{
$R_{\tau\to V}\equiv \Gamma (\tau^-\to\nu_\tau V^-) / \Gamma_{\tau\to e}$}
\\
\cline{2-3}    &  $\tau^-\to\nu_\tau V^-$ & $e^+e^-\to V^0$ 
\\ \hline 
$\pi^-\pi^0$  & $1.413\pm0.012$ & $1.360\pm0.043$ 
\\
$2\pi^-\pi^+\pi^0$ & $0.239\pm0.005$ & $0.239\pm0.031$ 
\\
$\pi^-3\pi^0$ & $0.064\pm0.008$ & $0.059\pm0.006$ 
\\
$\pi^-\omega$ & $0.108\pm0.004$ & $0.098\pm0.011$ 
\\
$3\pi^-2\pi^+\pi^0$ &	$0.0012\pm0.0003$ & $\geq 0.0010$
\\
$(6\pi)^-$		& \hfil --- \hfil  & $\geq 0.0052 $
\\
$\pi^-\pi^0\eta$ & $0.0101\pm0.0013$ & $0.0072\pm0.0011$ 
\\
$K^-K^0$ & $0.0089\pm0.0013$ & $0.0062\pm0.0016$   
\\
$\pi^-\phi$ & $< 0.002$ & $< 0.0006$
\\ \hline
\end{tabular}
\end{table}
%

The most recent results \cite{Eidelman} are compared
with the $\tau$--decay measurements in Table~\ref{tab:vector}.
The agreement is quite good.
Moreover, the experimental precision of the $\tau$--decay data is
already better than the $e^+e^-$ one. 

The exclusive $\tau$ decays into final hadronic states with $J^P = 1^+$,
or Cabibbo suppressed  modes  with $J^P =1^-$,
cannot  be  predicted  with  the  same degree  of
confidence. We can only make model--dependent estimates \cite{PI:89}
with an accuracy which
depends on our ability to handle the strong interactions at low energies.
That just indicates that the decay of the $\tau$ lepton is
providing new experimental hadronic information.  
Due to their semileptonic character, the hadronic $\tau$--decay data
are a unique and extremely useful tool to learn
about the couplings of the low--lying mesons to the weak currents.
 
\subsection{Chiral Dynamics}

At low momentum transfer, the coupling of any 
number of $\pi $'s, $K$'s and $\eta$'s to the
V$-$A current can be rigorously calculated with
Chiral Perturbation Theory techniques \cite{GL:85,chpt:95,EC:95}.
In the absence of quark masses the QCD Lagrangian splits into two
independent chirality (left/right) sectors, with their own quark
flavour symmetries.
With three light quarks ($u$, $d$, $s$), the QCD Lagrangian
is then approximately
invariant under chiral $SU(3)_L\otimes SU(3)_R$ rotations
in flavour space. The vacuum is however not symmetric under the
chiral group. Thus, 
chiral symmetry breaks down to the usual eightfold--way $SU(3)_V$,
generating the appearance of eight Goldstone bosons in the
hadronic spectrum, which can be identified with the lightest
pseudoscalar octet; their small masses being generated by the 
quark mass matrix, which explicitly breaks chiral symmetry.
The Goldstone nature of the pseudoscalar octet implies strong
constraints on their low--energy interactions, which can be worked out
through an expansion in powers of momenta over the 
chiral symmetry--breaking scale \cite{GL:85,chpt:95,EC:95}.

At lowest order in momenta, the 
couplings of the Goldstones to the weak current
can be calculated in a straightforward way.
The one--loop corrections are known \cite{GL:85,chpt:95,EC:95,CFU:96}
for the lowest--multiplicity
states ($\pi$, $K$, $2\pi$, $K\bar K$, $K\pi$, $3\pi$). 
Moreover, a two--loop calculation 
for the $2\pi$ decay mode is already available \cite{CFU:96}.
Therefore, exclusive hadronic $\tau$ decay data at low values of $q^2$
could be compared with rigorous QCD predictions.

There are also well--grounded theoretical results
(based on a $1/M_\rho$ expansion) for decays such as 
$\tau^-\to\nu_\tau(\rho\pi)^-,\nu_\tau(K^*\pi)^-,\nu_\tau(\omega\pi)^-$, 
but only in the
kinematical configuration where the pion is soft \cite{Wise}.

$\tau$ decays involve, however, high values of momentum transfer
where the chiral symmetry predictions no longer apply.
Since the relevant hadronic dynamics is governed by the non-perturbative
regime of QCD, we are unable at present to make first--principle
calculations for exclusive decays.
Nevertheless, one can still construct reasonable models, taking
into account the low--energy chiral theorems.
The simplest prescription \cite{PI:89,FWW:80,PI:87,GGP:90}
consist in extrapolating the chiral predictions
to higher values of $q^2$, by suitable final--state--interaction 
enhancements  which take into account the resonance structures
present in each channel in a phenomenological way.
This can be done weighting the contribution of a given set of
pseudoscalars, with definite quantum numbers, with an appropriate
resonance form factor. 
%
%
The requirement that the chiral predictions must be recovered
below the resonance region fixes the normalization of those form factors
to be one at zero invariant mass.

The extrapolation of the low--energy chiral theorems provides
a useful description of the $\tau$ data in terms of a few resonance
parameters. 
Therefore, it has been extensively used 
\cite{KS:90,PI:89,FWW:80,PI:87,GGP:90,karlsruhe,DE:94}
to analyze the
main $\tau$ decay modes, and has been incorporated into
the TAUOLA Monte Carlo library \cite{tauola}.
However, the model is too naive to be considered as
an actual implementation of the QCD dynamics.
Quite often, the numerical predictions could be drastically changed
by varying some free parameter or modifying the form--factor
ansatz.
Not surprisingly, some predictions fail badly to reproduce the
experimental data whenever a new resonance structure shows up
\cite{Shelkov}.

The addition of resonance form factors to the chiral low--energy
amplitudes does not guarantee that the
chiral symmetry constraints on the
resonance couplings have been correctly implemented. 
The proper way of including higher--mass states into the effective 
chiral theory was developed in Refs.~\cite{EGLPR:89}.
Using these techniques, a refined calculation of the rare decay
$\tau^-\to\nu_\tau\eta\pi^-$ has been given recently \cite{NR:95}.
A systematic analysis of $\tau$--decay amplitudes within this
framework is in progress \cite{GP:96}.

Tau decays offer a very good laboratory to improve our present
understanding of the low--energy QCD dynamics. 
The general form factors characterizing the non-perturbative
hadronic decay amplitudes
can be experimentally extracted from the
Dalitz--plot distributions of the final hadrons \cite{KM:92}.
An exhaustive analysis of $\tau$ decay modes
would provide a very valuable data basis to confront with
theoretical models.
 
\subsection{The Tau Hadronic Width}
\label{subsec:hadronic_width}

The inclusive character of the total $\tau$ hadronic width
renders possible an accurate calculation of the ratio
\cite{BR:88,NP:88,ORSAY:90,BNP:92,LDP:92a,OHIO:92,QCD:94,NA:95,Braaten}
%
\bel{eq:r_tau_def}
R_\tau \equiv { \Gamma [\tau^- \rightarrow \nu_\tau
                   \,\mbox{\rm hadrons}\, (\gamma)] \over
                         \Gamma [\tau^- \rightarrow
                \nu_\tau e^- {\bar \nu}_e (\gamma)] } ,
\ee
using analyticity constraints and the Operator Product Expansion
(OPE).

The theoretical analysis of $R_\tau$ involves
the two--point correlation functions 
\bel{eq:pi_j}
\Pi^{\mu\nu}_j(q)\equiv i \int d^4x \, e^{iqx} 
\langle 0|T(j^\mu(x) j^\nu(0)^\dagger)|0\rangle
\ee
for the vector, 
$j^\mu = V^{\mu}_{ij} \equiv \bar{\psi}_j \gamma^{\mu} \psi_i$,
and axial--vector,
$j^\mu = A^{\mu}_{ij} \equiv \bar{\psi}_j \gamma^{\mu} \gamma_5 \psi_i$,
colour--singlet quark currents ($i,j=u,d,s$).
They have the Lorentz decompositions
\beqn\label{eq:lorentz}
\Pi^{\mu \nu}_{ij,V/A}(q) & \!\!\!\! = & \!\!\!\!
  (-g^{\mu\nu} q^2 + q^{\mu} q^{\nu}) \, \Pi_{ij,V/A}^{(1)}(q^2)   \no\\
  && \!\!\!\! +   q^{\mu} q^{\nu} \, \Pi_{ij,V/A}^{(0)}(q^2) ,
\eeqn
where the superscript $(J=0,1)$ 
denotes the angular momentum in the hadronic rest frame.
  
The imaginary parts of the two--point functions
$\, \Pi^{(J)}_{ij,V/A}(q^2) \, $ 
are proportional to the spectral functions for hadrons with the 
corresponding quantum numbers.  The hadronic decay rate of the $\tau$
can be written as an integral of these spectral functions
over the invariant mass $s$ of the final--state hadrons:
\beqn\label{eq:spectral}
R_\tau  &\!\!\!\!\! = &\!\!\!\!\! 
12 \pi \int^{m_\tau^2}_0 {ds \over m_\tau^2 } \,
 \left(1-{s \over m_\tau^2}\right)^2 
\\ &\!\!\!\!\! \times &\!\!\!\!\! 
\biggl[ \left(1 + 2 {s \over m_\tau^2}\right) 
 \mbox{\rm Im} \Pi^{(1)}(s)
 + \mbox{\rm Im} \Pi^{(0)}(s) \biggr]  .\no 
\eeqn
 The appropriate combinations of correlators are 
\beqn\label{eq:pi}
\Pi^{(J)}(s)  &\!\!\! \equiv  &\!\!\!
  |V_{ud}|^2 \, \left( \Pi^{(J)}_{ud,V}(s) + \Pi^{(J)}_{ud,A}(s) \right)
\no\\ &\!\!\! + &\!\!\!
|V_{us}|^2 \, \left( \Pi^{(J)}_{us,V}(s) + \Pi^{(J)}_{us,A}(s) \right). 
\eeqn

We can separate the inclusive contributions associated with
specific quark currents:
\be\label{eq:r_tau_v,a,s}
 R_\tau \, = \, R_{\tau,V} + R_{\tau,A} + R_{\tau,S}\, .
\ee
$R_{\tau,V}$ and $R_{\tau,A}$ correspond to the first two terms
in \eqn{eq:pi}, while  
$R_{\tau,S}$ contains the remaining Cabibbo--suppressed contributions.
Non-strange hadronic decays of the $\tau$ are resolved experimentally
into vector ($R_{\tau,V}$) and axial-vector ($R_{\tau,A}$)
contributions according to whether the
hadronic final state includes an even or odd number of pions.
Strange decays ($R_{\tau,S}$) are of course identified by the
presence of an odd number of kaons in the final state.

\begin{figure}[tbh]
\label{fig:circle}
\centerline{\epsfxsize =7.5cm \epsfbox{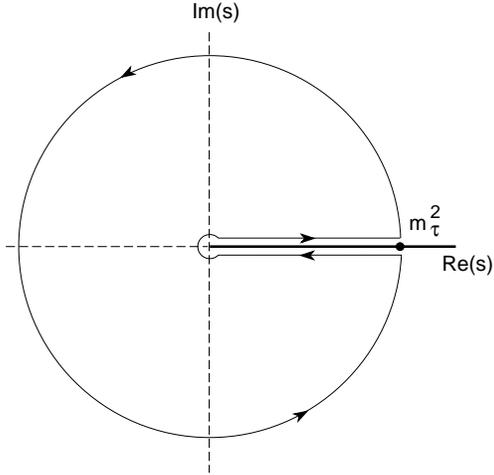}}
\vspace{-0.5cm}
\caption{Integration contour in the complex s--plane, used to obtain
Eq.~\protect\eqn{eq:circle}}
\end{figure}

Since the hadronic spectral functions are sensitive to the non-perturbative
effects of QCD that bind quarks into hadrons, the integrand in 
Eq.~\eqn{eq:spectral} cannot be calculated at present from QCD.
Nevertheless the integral itself can be calculated systematically
by exploiting
the analytic properties of the correlators $\Pi^{(J)}(s)$.
They are analytic
functions of $s$ except along the positive real $s$--axis, where their
imaginary parts have discontinuities.  
$R_\tau$ can
therefore be expressed as a contour integral 
in the complex $s$--plane running
counter-clockwise around the circle $|s|=m_\tau^2$:
\beqn\label{eq:circle}
 R_\tau &\!\!\!\!\! =&\!\!\!\!\! 
6 \pi i \oint_{|s|=m_\tau^2} {ds \over m_\tau^2} \,
 \left(1 - {s \over m_\tau^2}\right)^2
\\ &\!\!\!\!\!\times &\!\!\!\!\!
 \left[ \left(1 + 2 {s \over m_\tau^2}\right) \Pi^{(0+1)}(s)
         - 2 {s \over m_\tau^2} \Pi^{(0)}(s) \right]  . \no
\eeqn

The advantage of expression \eqn{eq:circle}  
over \eqn{eq:spectral}
is that it requires the correlators only for
complex $s$ of order $m_\tau^2$, which is significantly larger than 
the scale
associated with non-perturbative effects in QCD.  The short--distance
OPE can therefore be used to organize
the perturbative and non-perturbative contributions
to the correlators into a systematic expansion \cite{SVZ:79}
in powers of $1/s$.
 The possible uncertainties associated with the use of the OPE near the
 time-like axis are negligible in this case, because
 the integrand in \eqn{eq:circle} includes a factor
 $(1- s/m_\tau^2)^2$, which provides a double zero at $s=m_\tau^2$,
 effectively suppressing the contribution from the
 region near the branch cut.

After evaluating the contour integral, $R_\tau$
can be expressed as an expansion in powers of $1/m_\tau^2$,
with coefficients that depend only logarithmically on $m_\tau$:
\bel{eq:r_total}
R_{\tau} = 3
S_{EW} \left\{ 1 + \delta_{EW}'  
+ \sum_{D=0,2,...} \delta^{(D)}\right\} .
\ee
The factors $S_{EW}=1.0194$ and $\delta_{EW}'=0.0010$ 
contain the known electroweak corrections at the leading
\cite{MS:88} and next-to-leading \cite{BL:90} logarithm
approximation.
The dimension--0 contribution,
$\delta^{(0)}$,
is the purely perturbative correction neglecting quark masses.
It is given by
\cite{BR:88,NP:88,ORSAY:90,BNP:92,LDP:92a,OHIO:92,QCD:94}:
\beqn\label{eq:delta0}
\delta^{(0)}  &\!\!\! =&\!\!\! \sum_{n=1}  K_n \, A^{(n)}(\alpha_s) 
\no\\ &\!\!\! = &\!\!\!
a_\tau + 5.2023 a_\tau^2 + 26.366 a_\tau^3
       + \, \cO(\alpha_s^4)  \, ,
\eeqn
where $a_\tau\equiv\alpha_s(m_\tau^2)/\pi$.

The dynamical coefficients $K_n$ regulate the perturbative expansion
of  $ -s{d\over ds}\Pi^{(0+1)}(s)$ 
in the massless--quark limit
[$s\Pi^{(0)}(s)=0$ for massless quarks]; they are
known \cite{ChKT:79,GKL:91,SS:91} to $\cO(\alpha_s^3)$:
$K_1 = 1$; $K_2 = 1.6398$; $K_3(\overline{MS}) = 6.3711$.
The kinematical effect of the contour integration is contained in
the functions \cite{LDP:92a}
\beqn\label{eq:a_xi}
A^{(n)}(\alpha_s) &\!\!\!\! = &\!\!\!\! {1\over 2 \pi i}
\oint_{|s| = m_\tau^2} {ds \over s} \,
  \left({\alpha_s(-s)\over\pi}\right)^n  
\no\\ &\!\!\!\!\times  &\!\!\!\! 
 \left( 1 - 2 {s \over m_\tau^2} + 2 {s^3 \over m_\tau^6}
         - {s^4 \over  m_\tau^8} \right) , 
\eeqn
which only depend on $\alpha_s(m_\tau^2)$.
Owing to the long running of the strong coupling along the circle, the
coefficients of the
perturbative expansion of $\delta^{(0)}$ in powers of
$\alpha_s(m_\tau^2)$ are larger than the direct $K_n$
contributions. This running effect can be properly resummed to all orders
in $\alpha_s$ by fully keeping \cite{LDP:92a}
the known three--loop--level calculation of
the integrals $A^{(n)}(\alpha_s)$.

The leading quark--mass corrections $\delta^{(2)}$ are known
\cite{BNP:92,QCD:94,CK:93} to order $\alpha_s^2$.
They are certainly tiny for the up and down quarks
($\delta^{(2)}_{ud}\sim -0.08\% $), but 
the correction from the strange quark mass is important
for strange decays
($\delta^{(2)}_{us}\approx  -19\% $).
Nevertheless, because of the $|V_{us}|^2$ suppression, the
effect on the total ratio $R_{\tau}$  is only $-(0.9\pm 0.2) \%$.

The leading non-perturbative contributions can be shown to be
suppressed by six powers of the $\tau$ mass
\cite{BR:88,NP:88,ORSAY:90,BNP:92},
and are therefore very small.
This fortunate fact is due to the phase--space factors in
\eqn{eq:circle}; their form is such that the leading $1/s^2$ corrections
to $\Pi^{(1)}(s)$ do not survive the integration along the circle.

The numerical size of the non-perturbative corrections can be 
determined from the
invariant--mass distribution of the final hadrons in $\tau$ decay \cite{PI:89}.
Although the distributions themselves cannot be predicted at present, certain
weighted integrals of the hadronic spectral functions can be calculated in the
same way as $R_\tau$, and used to extract the non-perturbative contributions
from the data themselves \cite{PI:89,LDP:92b}.
The predicted suppression \cite{BR:88,NP:88,ORSAY:90,BNP:92}
of the non-perturbative corrections has been confirmed by
ALEPH \cite{ALEPH:93} and CLEO \cite{CLEO:95}.
The most recent ALEPH analysis \cite{Hocker} gives:
\bel{eq:del_np}
\delta_{\mbox{\rms NP}} \equiv \sum_{D\geq 4} \delta^{(D)} =
(0.5\pm 1.1)\% \, ,
\ee
in agreement with previous estimates \cite{BNP:92}.

%
\begin{table}[tbh]
\centering
\caption{$\delta^{(0)}$ for different values of
$\overline\alpha_s\equiv\alpha_s(m_\tau^2)$ \protect\cite{QCD:94}}
\label{tab:perturbative}
\vspace{0.2cm}
\begin{tabular}{cc|cc}  
\hline  
$\overline\alpha_s$ & $\delta^{(0)}$ &
$\overline\alpha_s$ & $\delta^{(0)}$  
\\ \hline
%
$0.24$ & $0.118\pm0.003$   &
$0.34$ & $0.191\pm0.009$
\\
$0.26$ & $0.132\pm0.004$   &   
$0.36$ & $0.205\pm0.010$
\\       
$0.28$ & $0.146\pm0.005$   &
$0.38$ & $0.220\pm0.012$
\\
$0.30$ & $0.161\pm0.006$   &
$0.40$ & $0.234\pm0.013$
\\
$0.32$ & $0.176\pm0.008$   &
$0.42$ & $0.248\pm0.013$
\\  \hline   
\end{tabular}
\end{table}
%

The QCD prediction
for $R_\tau$ is then completely dominated by the
perturbative contribution $\delta^{(0)}$; non-perturbative effects being smaller
than the perturbative uncertainties from uncalculated higher--order
corrections \cite{QCD:94,NA:95,Braaten,Raczka}. Furthermore, 
as shown in Table~\ref{tab:perturbative}, the result turns out to be
very sensitive to the value of $\alpha_s(m_\tau^2)$, allowing for an accurate
determination of the fundamental QCD coupling.

The experimental value for $R_\tau$ can be obtained
from the leptonic branching fractions or from the
$\tau$ lifetime. The average of those determinations
\be  
  R_\tau =3.649 \pm 0.014 \, ,
\ee
corresponds to 
\be\label{eq:alpha}
\alpha_s(m_\tau^2)  =  0.35\pm0.03 \, . 
\ee

Once the running coupling constant $\alpha_s(s)$ is determined at the scale
$m_\tau$, it can be evolved to higher energies using the renormalization
group.  The size of its error bar scales roughly
as $\alpha_s^2$, and it therefore shrinks as the scale increases.
Thus a modest precision in the
determination of $\alpha_s$ at low energies results in a very high
precision in the coupling constant at high energies.
After evolution up to the scale $M_Z$, the strong coupling constant in
\eqn{eq:alpha} decreases to\footnote{
From a combined analysis of $\tau$ data, ALEPH quotes
\protect\cite{Hocker}:
$\alpha_s(M_Z^2)  =
0.1225\pm 0.0006_{\mbox{\protect\rms exp}}\pm
0.0015_{\mbox{\protect\rms th}}\pm 0.0010_{\mbox{\protect\rms evol}}$.  
}
%
\be\label{eq:alpha_z}
\alpha_s(M_Z^2)  =  0.122\pm 0.003 \, ,
\ee
in excellent
agreement with the present LEP average \cite{BE:96} (without $R_\tau$)
$\alpha_s(M_Z^2)  =  0.122\pm0.006$  
and with a smaller error bar.
The comparison of these two determinations of $\alpha_s$ in two extreme
energy regimes, $m_\tau$ and $M_Z$, provides a beautiful test of the
predicted running of the QCD coupling.

Using the measured invariant--mass distribution of the final hadrons, it
is possible to evaluate the integral \eqn{eq:spectral}, with an arbitrary
upper limit of integration $s_0\leq m_\tau$. The experimental $s_0$
dependence  agrees well with the
theoretical predictions \cite{LDP:92b} up to rather low values of $s_0$. 
Equivalently,  from the measured $R_\tau(s_0)$ distribution one obtains
$\alpha_s(s_0)$ as a function of the scale $s_0$ in good agreement with the
running predicted at three--loop order by QCD \cite{GN:96}.

With $\alpha_s(m_\tau^2)$ fixed to the value in Eq.~(\ref{eq:alpha}), 
the same theoretical framework gives definite
predictions \cite{BNP:92,QCD:94} for the semi-inclusive $\tau$ decay widths
$R_{\tau,V}$, $R_{\tau,A}$ and $R_{\tau,S}$, in good agreement with the
experimental measurements \cite{Hocker,Davier}.
The analysis of these semi-inclusive quantities (and the
associated invariant--mass distributions \cite{LDP:92b})
provides important information on several QCD parameters.
For instance, $R_{\tau,V}-R_{\tau,A}$ is a pure non-perturbative
quantity; 
basic QCD properties force
the associated invariant--mass distribution to obey a
series of chiral sum rules \cite{PI:89,Hocker}.
The Cabibbo--suppressed width $R_{\tau,S}$ is very
sensitive to the value of the strange quark mass \cite{BNP:92},
providing a direct and clean way of measuring $m_s$;
a very preliminary value has been already presented at this
workshop \cite{Davier}.
Last but not least, the measurement of the vector spectral function 
\cite{Alemany}
Im$\Pi_V(s)$ helps to reduce the present uncertainties in fundamental
QED quantities such as $\alpha(M_Z)$ and $(g-2)_\mu$.

\section{SEARCHING FOR NEW PHYSICS}
\label{sec:new-physics}

\subsection{Lepton--Number Violation}

In the minimal SM with massless neutrinos, there is a
separately conserved additive lepton number for each generation. All
present data are
consistent with this conservation law. However, there are no strong
theoretical
reasons forbidding a mixing among the  different  leptons, in the same
way as happens in the quark sector.
Many models in fact predict lepton--flavour or even
lepton--number violation at some level.
Experimental searches for these processes
can provide information on the scale at which the new physics begins to
play a  significant role.

$K$, $\pi$ and $\mu$ decays, together with $\mu$--$e$ conversion,
neutrinoless double beta
decays  and  neutrino  oscillation  studies,  have  put  already
stringent  limits \cite{PDG:96} on
lepton--flavour and lepton--number violating interactions.
However, given the present lack of
understanding of the origin of fermion generations, one can imagine
different
patterns of violation of this conservation law  for  different  mass
scales.
Moreover, the larger mass of the $\tau$ opens the possibility of 
new types of decay which are kinematically forbidden for the $\mu$.
 
The present upper limits on lepton--flavour and
lepton--number violating decays of the $\tau$ \cite{PDG:96,Gan} are 
in the range of $10^{-4}$ to $10^{-6}$, which is far
away from the impressive bounds \cite{PDG:96} obtained in $\mu$ decay
[$Br(\mu^-\to e^- \gamma)  < 4.9 \times 10^{-11},
Br(\mu^-\to e^- e^+ e^-) < 1.0 \times 10^{-12},
Br(\mu^-\to e^-\gamma\gamma) <  7.2 \times 10^{-11} \, $ (90\% CL)].
With future $\tau$--decay samples of $10^7$ events
per year, an improvement of two orders of magnitude would be possible.
 
The lepton--flavour violating couplings of the
$Z$ boson have been investigated at LEP. The present ($95\% $ CL)
limits are \cite{OPAL:95c}:
\beqn\label{eq:Z_LFV}
\mbox{\rm Br}(Z\to e^\pm\mu^\mp) &<& 1.7 \times 10^{-6} ; \no\\
\mbox{\rm Br}(Z\to e^\pm\tau^\mp) &<& 9.8 \times 10^{-6}  ; \\
\mbox{\rm Br}(Z\to \mu^\pm\tau^\mp) &<& 1.7 \times 10^{-5}  .\no
\eeqn

\subsection{The Tau Neutrino}

All observed $\tau$ decays are supposed to be accompanied by neutrino
emission, in order to fulfil energy--momentum conservation requirements.
The present data are consistent with the $\nu_\tau$
being a conventional sequential neutrino. Since taus are not produced
by $\nu_e$ or $\nu_\mu$ beams, we know that $\nu_\tau$
is different from the electronic and  muonic
neutrinos, and a (90\% CL) upper limit can be set on the couplings of the 
$\tau$ to
$\nu_e$ and $\nu_\mu$ \cite{E531:86}:
\bel{eq:taunulimits}
|g_{\tau\nu_e}| < 0.073 \ , \qquad |g_{\tau\nu_\mu}| <
        0.002 \ .  
\ee
These limits can be interpreted in terms of $\nu_e/\nu_\mu\to\nu_\tau$
oscillations, to exclude a
region in the neutrino mass--difference and neutrino mixing--angle space.
In the extreme situations of large $\delta m^2$ or maximal mixing,
the limits are \cite{E531:86}:
\beqn\label{eq:numu_nutau}
\lefteqn{\nu_\mu\to\nu_\tau :}
\no\\ &&
\sin^2{2\theta_{\mu,\tau}} < 0.004 \qquad 
    (\mbox{\rm large}\, \delta m_{\mu,\tau}^2) , 
\no\\ &&
\delta m_{\mu,\tau}^2 < 0.9 \,\mbox{\rm eV}^2 \quad\;\; 
   (\sin^2{2\theta_{\mu,\tau}} = 1) ;
\\ \label{eq:nue_nutau}
\lefteqn{\nu_e\to\nu_\tau : }
\no\\ &&
\sin^2{2\theta_{e,\tau}} < 0.12 \qquad\;\;\,
    (\mbox{\rm large}\, \delta m_{e,\tau}^2) , 
\no \\ &&
\delta m_{e,\tau}^2 < 9 \,\mbox{\rm eV}^2 \qquad\;\,\, 
   (\sin^2{2\theta_{e,\tau}} = 1) .
\eeqn
The new CHORUS \cite{Gregoire} and NOMAD \cite{Cardini} experiments,
presently running at CERN, and the future Fermilab E803 experiment 
are expected to improve the
$\nu_\mu\to\nu_\tau$  oscillation limits 
by at least an order of magnitude.

LEP and SLC have confirmed \cite{LEP:96}
the existence of three (and only
three) different light neutrinos, with standard couplings to the $Z$.
However,
no direct observation of $\nu_\tau$, that is, interactions resulting
from neutrinos produced in $\tau$ decay, has been made so far.

The expected source of tau neutrinos in beam dump
experiments is
the decay of $D_s$ mesons produced by interactions in the dump; i.e.,
$p+N\to D_s + \cdots $
followed by the decays $D_s\to\tau^-\bar\nu_\tau$
and $\tau^-\to\nu_\tau + \cdots $\  
Several experiments \cite{BEBC:87}
have searched for
$\, \nu_\tau + N \to \tau^- + \cdots $\
interactions with negative results; therefore, only an upper
limit on the production of $\nu_\tau$'s has been obtained.
The direct detection of the $\nu_\tau$
should be possible \cite{dRR:84} at the LHC,
thanks to the large charm production cross-section of this collider.

The possibility of a non-zero neutrino mass is obviously a
very important
question in particle physics \cite{Langacker}. 
There is no fundamental principle requiring
a null mass for the neutrino. On the contrary, many extensions of the SM
predict non-vanishing neutrino masses, which could have, in addition,
important implications in cosmology and astrophysics.
The strongest bound up to date is the preliminary ALEPH limit
\cite{Passalacqua}, 
\be\label{eq:numasslimit}
m_{\nu_\tau} \, < \, 18.2 \,\mbox{\rm MeV} \quad (95\%\,
\mbox{\rm CL}),
\ee
obtained from a two--dimensional likelihood fit of the
visible energy and the invariant--mass distribution of
$\tau^-\to (3 \pi)^-\nu_\tau, (5 \pi)^-\nu_\tau$ 
events. 
 
For comparison, the present limits on the muon
and electron
neutrinos are \cite{PDG:96}
$m_{\nu_\mu} < 170$ KeV    (90\% C.L.)
and $m_{\nu_e} < 15$ eV.   
Note, however, that in many models a mass hierarchy among
different generations is expected, with the neutrino mass being
proportional to some power of the mass of its charged lepton partner.
Assuming for instance the  fashionable  relation
$\, m_{\nu_\tau} / m_{\nu_e} \sim (m_\tau/m_e)^2$,
the bound \eqn{eq:numasslimit} would be equivalent to
a limit of 1.5 eV for $m_{\nu_e}$.
A relatively crude measurement of $m_{\nu_\tau}$
may then imply strong constraints on neutrino--mass model building.

More stringent (but model--dependent) bounds on $m_{\nu_\tau}$ can be
obtained from cosmological considerations.
A stable neutrino (or an unstable one with a lifetime comparable
to or longer than the age of the Universe) must not overclose the
Universe. Therefore, measurements of the age of the
Universe exclude stable neutrinos in  the range \cite{GZ:72,LW:77}
200 eV $< m_\nu <$ 2 GeV.
Unstable neutrinos with lifetimes longer than 300 sec
could increase the expansion rate of the Universe, 
spoiling the successful predictions for the
primordial nucleosynthesis of light isotopes in the early universe 
\cite{KO:91}; the mass range
0.5 MeV $< m_{\nu_\tau} < $ 30 MeV
has been excluded in that case \cite{KO:91,DR:93,KA:94,DGT:94,GT:95}. 
For neutrinos of any lifetime decaying into electromagnetic daughter
products, it is possible to exclude the same mass range, combining the
nucleosynthesis constraints with limits based on the supernova SN 1987A
and on BEBC data \cite{DGT:94,GT:95}.
Light neutrinos ($m_{\nu_\tau}< 100$ keV)
decaying through $\nu_\tau\to\nu_\mu + G^0$, 
are also excluded by the nucleosynthesis constraints, if their lifetime
is shorter than $10^{-2}$ sec \cite{KA:94}.

The astrophysical and cosmological arguments lead indeed to quite
stringent limits; however, they always involve (plausible) assumptions
which could be relaxed in some physical scenarios \cite{MS:95,FKO:96}.
For instance, in deriving the abundance of massive $\nu_\tau$'s at 
nucleosynthesis, it is always assumed that tau neutrinos annihilate at
the rate predicted by the SM.
A $\nu_\tau$ mass in the few MeV range (i.e. the mass sensitivity
which can be achieved in the foreseeable future) could have a host
of interesting astrophysical and cosmological consequences \cite{GT:95}:
relaxing the big-bang nucleosynthesis bound to the baryon density and
the number of neutrino species; allowing big-bang nucleosynthesis to
accommodate a low ($< 20\% $) ${}^4$He mass fraction or high ($>10^{-4}$)
deuterium abundance; improving significantly the agreement between the
cold dark matter theory of structure formation and observations
\cite{DGT:94b};
and helping to explain how type II supernova explode.

  The electromagnetic structure of the $\nu_\tau$ can be tested through
the process $e^+e^-\to\nu_\tau\bar\nu_\tau\gamma$. The combined data
from PEP and PETRA implies \cite{GR:88} the following $90 \% $ CL upper
bounds on the magnetic moment and charge radius  of the $\nu_\tau$
($\mu_B \equiv e\hbar / 2 m_e$):
$|\mu(\nu_\tau)|  <  4 \times 10^{-6} \, \mu_B$;
$<r^2>(\nu_\tau) <  2 \times 10^{-31} \, \mbox{\rm cm}^2$.
A better limit on the $\nu_\tau$ magnetic moment,
\bel{eq:mu_nu_tau}
|\mu(\nu_\tau)|  <  5.4 \times 10^{-7} \, \mu_B \quad (90\%\,\mbox{\rm CL}) ,
\ee
has been placed by the BEBC experiment \cite{BEBC:92},
by searching for elastic $\nu_\tau e$ scattering events, using a
neutrino beam from a beam dump which has a small $\nu_\tau$ component.

A big $\nu_\tau$ magnetic moment of about $10^{-6} \mu_B$ has been
suggested, in order to make the $\tau$ neutrino an acceptable cold
dark matter candidate. For this to be the case, however, the
$\nu_\tau$ mass should be in the range
1 MeV $ < m_{\nu_\tau} < $ 35 MeV  \cite{GI:90}.
The same region of $m_{\nu_\tau}$ has been suggested in trying to
understand the baryon--antibaryon asymmetry of the universe \cite{CKN:91}.

\subsection{Dipole Moments}

Owing to their chiral changing structure, the 
electroweak dipole moments 
may provide important insights on the mechanism responsible for
mass generation. In general, one expects \cite{MA:94}
that a fermion of mass
$m_f$ (generated by physics at some scale $M\gg m_f$) will have
induced  dipole moments proportional to some power of 
$m_f/M$.
Therefore, heavy fermions such as the $\tau$ should be a good 
testing ground for this kind of effects.
Of special interest are the electric and weak
dipole moments, $d^{\gamma,Z}_\tau$, 
which  violate $T$ and $P$ invariance; they constitute
a good probe of CP violation.

The more stringent (95\% CL) limits
on the anomalous magnetic moment and the electric dipole moment
of the $\tau$ have
been derived from an
analysis of the $Z\to\tau^+\tau^-$ decay width 
\cite{EM:93},
assuming that all other couplings take their SM values:
\beqn\label{eq:a_g_tau}
-0.004 < a^\gamma_\tau <0.006 \ , \no\\
|d^\gamma_\tau| < 2.7 \times 10^{-17} \,  e \, \mbox{\rm cm}.
\eeqn
These limits would be invalidated in the presence of any
CP--conserving contribution to $\Gamma(Z\to\tau^+\tau^-)$
interfering destructively with the SM amplitude.

Slightly weaker bounds
have been extracted from the decay 
$Z\to\tau^+\tau^-\gamma$  \cite{Wermes} (95\% CL):
\bel{eq:a_g_tau_direct}
|a^\gamma_\tau|  < 0.0104\ , \qquad 
|d^\gamma_\tau| < 5.8 \times 10^{-17} \,  e \,\mbox{\rm cm} ,
\ee
and from PEP and PETRA data \cite{SI:83,MA:89,AS:90}:
$|a^\gamma_\tau|(35\,\mbox{\rm GeV})  < 0.023$ (95\% CL),
$|d^\gamma_\tau|(35\,\mbox{\rm GeV}) < 1.6 \times 10^{-16} \,  e$ cm
(90\% CL). 

In the SM, $|d^\gamma_\tau|$ vanishes, while 
the overall value of $a^\gamma_\tau$
is dominated by the  second order QED contribution \cite{SC:48},
$a^\gamma_\tau \approx \alpha / 2 \pi$.
Including QED corrections up to O($\alpha^3$),
hadronic vacuum polarization contributions
and the corrections due to the weak interactions 
(which are a factor 380
larger than for the muon), the tau anomalous magnetic moment has been
estimated to be \cite{NA:78,SLM:91}
\bel{eq:a_th_tau}
a^\gamma_\tau\big |_{\mbox{\rms th}} \, = \, (1.1773 \pm 0.0003)
     \times 10^{-3} \, .
\ee

The first direct limit on the weak anomalous magnetic moment
has been obtained by L3, by using correlated azimuthal
asymmetries of the $\tau^+\tau^-$ decay products \cite{BGV:94}. 
The
preliminary (95\% CL) result of this analysis is \cite{ESanchez}:
\bel{eq:a_Z}
 -0.016 < a^Z_\tau < 0.011 \ .
\ee

The possibility of a CP--violating weak dipole moment of the $\tau$ has
been investigated at LEP, by studying
$T$--odd triple correlations \cite{BN:89,BE:91}
of the final $\tau$--decay products in $Z\to\tau^+\tau^-$ events.
The present (95\% CL) limits are \cite{Wermes}:
\beqn\label{eq:d_Z_tau}
\vert\mbox{\rm Re}\, d^Z_\tau(M_Z^2)\vert \le 3.6\times 10^{-18}
   \, e \,  \mbox{\rm cm} ,
\no\\
\vert\mbox{\rm Im}\, d^Z_\tau(M_Z^2)\vert \le 1.1\times 10^{-17}
   \, e \,  \mbox{\rm cm} .
\eeqn
These limits provide useful constraints on different models
of CP violation \cite{BN:89,BLMN:89,DR:91,KOR:91}.

T--odd signals can be also generated through
a relative phase between the vector and axial-vector couplings
of the $Z$ to the $\tau^+\tau^-$ pair \cite{BPR:91}, i.e.
$\mbox{\rm Im}(v_\tau a_\tau^*) \not= 0$.
This effect, which in the  SM appears \cite{BR:89} at the
one-loop level through absorptive parts in the electroweak amplitudes,
gives rise \cite{BPR:91} to a spin--spin correlation associated with the
transverse (within the production plane) and normal (to the production
plane) polarization components of the two $\tau$'s.
A preliminary measurement of these transverse spin correlations has been
reported by  ALEPH \cite{FSanchez}.

\subsection{CP Violation}

In the three--generation SM, the violation of the CP symmetry 
originates from the single phase 
naturally occurring in the quark mixing matrix \cite{KM:73} . 
Therefore, CP violation
is predicted to be absent in the lepton sector (for massless neutrinos).
The present experimental observations are in agreement with the
SM; nevertheless, the correctness of the
Kobayashi---Maskawa mechanism is far from being proved.
Like fermion masses and quark mixing angles, the origin of the
Kobayashi---Maskawa phase lies in the most obscure part of the
SM Lagrangian: the scalar sector.
Obviously, CP violation could well be a sensitive probe for new
physics.

Up to now, CP violation in the lepton sector
has been investigated mainly through the electroweak
dipole moments.
Violations of the CP symmetry could also happen in the
$\tau$ decay amplitude.
In fact, the possible CP--violating effects can be expected to be larger 
in $\tau$ decay than in $\tau^+\tau^-$ production \cite{tsai}. 
Since the decay of the $\tau$ proceeds through a weak interaction,
these effects could be ${\cal O}(1)$ or ${\cal O}(10^{-3})$,
if the leptonic CP violation is {\it weak} or {\it milliweak}
 \cite{tsai}.

With polarized electron (and/or positron) beams, one could use the
longitudinal polarization vectors of the incident leptons to construct
T--odd rotationally invariant products. CP could be tested by comparing
these T--odd products in $\tau^-$ and $\tau^+$ decays.
In the absence of beam polarization, CP violation could still be tested
through $\tau^+\tau^-$ correlations. In order to separate possible 
CP--odd effects in the $\tau^+\tau^-$ production and in the $\tau$ decay,
it has been suggested to study the final decays of the $\tau$--decay
products and build the so-called 
{\it stage--two spin--correlation functions} \cite{stscf}.
For instance, one could study the chain process
$e^+e^-\to\tau^+\tau^-\to (\rho^+\bar\nu_\tau)(\rho^-\nu_\tau)\to
(\pi^+\pi^0\bar\nu_\tau)(\pi^-\pi^0\nu_\tau)$. The distribution of the
final pions provides information on the $\rho$ polarization, which allows
to test for possible CP--violating effects in the 
$\tau\to\rho\nu_\tau$ decay.

CP violation could also be tested
through rate asymmetries, i.e. comparing the partial fractions
$\Gamma(\tau^-\to X^-)$ and $\Gamma(\tau^+\to X^+)$. However, this kind
of signal requires the presence of strong final--state interactions in the
decay amplitude.
Another possibility would be to study T--odd (CPT--even) asymmetries in the
angular distributions of the final hadrons in semileptonic 
$\tau$ decays \cite{FM:96}. 
Explicit studies of the decay modes
$\tau^-\to K^-\pi^-\pi^+, \pi^- K^- K^+$ \cite{KKSW:94} and
$\tau^-\to \pi^-\pi^-\pi^+$ \cite{ChHT:95} show that
sizeable CP--violating effects
could be generated in some models of CP violation 
involving several Higgs doublets
or left--right symmetry.

\section{SUMMARY}
\label{sec:summary}

The flavour structure of the SM is one of the main pending questions
in our understanding of weak interactions. Although we do not know the
reason of the observed family replication, we have learned experimentally
that the number of SM fermion generations is just three (and no more).
Therefore, we must study as precisely as possible the few existing flavours
to get some hints on the dynamics responsible for their observed structure.

The $\tau$ turns out to be an ideal laboratory to test the SM. 
It is a lepton, which means clean physics, and moreover it is
heavy enough to produce a large variety of decay modes.
Na\"{\i}vely, one would expect the $\tau$ to be much more sensitive
than the $e$ or the $\mu$ to new physics related to the flavour and
mass--generation problems.

QCD studies can also benefit a lot from the existence of this heavy lepton,
able to decay into hadrons. Owing to their semileptonic character, the
hadronic $\tau$ decays provide a powerful tool to investigate the low--energy
effects of the strong interactions in rather simple conditions.

Our knowledge of the $\tau$ properties has been considerably
improved during the last few years. 
Lepton universality has been tested to rather good accuracy,
both in the charged and neutral current sectors. The
Lorentz structure of the leptonic $\tau$ decays is certainly not determined,
but begins to be experimentally explored.  The quality of the hadronic data
has made possible to perform quantitative QCD tests
and determine the strong coupling constant very accurately.
Searches for non-standard phenomena have been pushed to the limits 
that the existing data samples allow to investigate.

At present, all experimental results on the $\tau$ lepton are consistent with
the SM. There is, however, large room for improvements. Future $\tau$ experiments
will probe the SM to a much deeper level of sensitivity and will explore the
frontier of its possible extensions.

\section*{ACKNOWLEDGEMENTS}
I would like to thank the organizers for creating a very stimulating atmosphere.
Useful discussions with Ricard Alemany, Michel Davier and Andreas H\"ocker
are also acknowledged.
I am indebted to Manel Martinez for keeping me informed about the
most recent LEP averages, and to Wolfgang Lohmann for providing the PAW files to
generate figures 3 and 4.
This work has been supported in part by CICYT (Spain) under grant 
No. AEN-96-1718.

\end{document}